\documentclass{iopart}
\usepackage{graphicx}
\begin{document}
\title[]{\center{Leptogenesis and Low energy CP violation, a link}}
\author{T. Endoh\dag, S. Kaneko\ddag, S. K. Kang\dag\S, \\
T. Morozumi\footnote[1]{Talk presented by T. Morozumi,
morozumi@hiroshima-u.ac.jp} and M. Tanimoto\ddag
}
\address{\dag\ Graduate School of Science, Hiroshima University, 739-8526, 
Japan}
\address{\ddag\ Department of Physics, Niigata University,
Niigata, 950-2181, Japan}
\address{\S\ Department of Physics, Seoul National University,
Seoul, 151-747, Korea}
\def\bea{\begin{eqnarray}}
\def\eea{\end{eqnarray}}
\def\nn{\nonumber}
\begin{abstract}
How is CP violation of low energy related to  CP violation required from baryon number asymmetry ?  We give an example which shows a direct link 
between CP violation of neutrino oscillation and baryogenesis 
through leptogenesis.
\end{abstract}
When the sphaleron process is active, the sum of baryon number and
lepton number is not a conserved quantity;
 $ \frac{d(B+L)}{dt}=$ Anomaly $\ne 0 $. Therefore, the evolution equation 
 of baryon number and lepton number becomes a coupled equation. The present 
 baryon number can be written in terms of the "initial" lepton number and 
 baryon number as;
 \bea
B_{now}=\frac{1}{2}(B-L)_{ini.}+ \frac{1}{2}(B+L)_{ini.} 
\exp[-\frac{\Delta t}{\tau}]
\rightarrow  \frac{1}{2}(B-L)_{ini.}.
\eea
Fukugita and Yanagida proposed "baryogenesis without Grand unification"
\cite{FuYa}. Then, $B_{ini.}=0$ while the lepton number production is possible
because their model includes the heavy Majorana neutrinos and their decays
lead to the lepton number asymmetry;
\bea
L_{ini} \sim
\Gamma[N  \rightarrow l^- \phi^+] -\Gamma[N \rightarrow l^+ \phi^-]. 
\eea
The purpose of my talk is give a specific example which shows 
"a direct link"
between  the size and sign of baryon number  and CP violation in neutrino
oscillatio; 
\bea
P(\nu_{\mu} \rightarrow \nu_{e})-P(\bar{\nu_\mu}
\rightarrow \bar{\nu_e}) \sim J
=Im (K_{e1} K_{e2}^{\ast})(K_{\mu 1} K_{\mu 2}^{\ast})^{\ast},
\label{J}
\eea
where  $K$ is MNS matrix.
\bea
J_{\mu}^{cc}= \bar{l}_L \gamma_{\mu} K \nu_L.  
\eea
$J \sim P-\bar{P}=\Delta P$ 
is related to MNS matrix. By taking the basis in which the
mass matrix for the 
heavy Majorana neutrinos  and charged leptons are real diagonal,
the MNS matrix can be obtained through the diagonalization of
$ m_{eff}=-m_D \frac{1}{M} {m_D}^T.$ 
On the other hand the lepton number asymmetry in the same basis is given as;
\bea
\epsilon_1 &=& 
\frac{\Gamma[N_1 \rightarrow l^- \phi^+]-\Gamma[N_1 \rightarrow l^+ \phi^-]}
{\Gamma[N_1 \rightarrow l^- \phi^+]+\Gamma[N_1 \rightarrow l^+ \phi^-]}
=
-\frac{3 M_1}{2 M_2}
\frac{Im [(m_D^{\dagger} m_D)_{12}^2]}{V ^2 (m_D^{\dagger} m_D)_{11}}, 
\eea
where $V \sim 1$ (TeV) and we take $M_2 \gg M_1$.
We consider "the minimal seesaw" which generates $L \ne 0$ and
$\Delta P \ne 0$ simultaneously. The minimal model is
$(3,2)$ model $ with $ 3 light neutrinos  $\nu_1, \nu_2, \nu_3$ and
2 heavy Majorana neutrionos $N_1$ and $N_2$.
\bea
{\cal L}=&& \bar{l^i} {m_l}_i l^i + \bar{\nu^i} {m_D}_{ij} N_{Rj} +\frac{1}{2}
\bar{(N_{Rj})^c} M_j N_{Rj},
\eea
where $ i=1 \sim 3, j=1,2$.  
Dirac mass term $3 \times 2$ matrix,
\bea
m_D&=&\left(\begin{array}{cc}
      m_{11} & m_{12}\\
      m_{21} & m_{22}\\
      m_{31} & m_{32}\\ 
            \end{array} \right) 
   =U_L \left(\begin{array}{cc}
      0 & 0 \\
      m_2 & 0 \\
      0 & m_3  \\ 
            \end{array} \right) V_R,
\eea
where $U_L$ ($V_R$) is a $3 \times 3$ ($ 2 \times 2$) unitary matrix.
The important properties of the model are
 one light neutrino is exactly massless:
$ det [m_D \frac{1}{M} m_D ^T]=0$
and there are three  CP violating phases since there are
$3 =6-3 $ imaginary elements in $m_D$.  By writing $V_R$ as follows;
\bea
V_R&=&\left(\begin{array}{cc}
      c_R  & s_R  \\
      -s_R  & c_R \end{array} \right)
      \left(\begin{array}{cc}
      \exp(-i {\frac{\gamma_R}{2}})  & 0 \\
      0 & \exp(i {\frac{\gamma_R}{2}}) 
      \end{array} \right),
\eea
we can show 
the leptogenesis is determined  by a CP phase {$\gamma_R$},
\bea
\epsilon_1 && \sim - Im[(m_D^{\dagger} m_D)_{12}^2]  \nn \\
    && \sim -({m_3}^2-{m_2}^2)^2 
{s_R}^2 {c_R}^2 \sin {2 \gamma_R}.
\eea
On the other hand, CP violation in neutrino oscillation, $J$ in Eq.(\ref{J}),
depends on  all three CP violating phases because 
$K$ is determined by the diagonalization as
$-K^{\dagger} m_D \frac{1}{M} m_D^T  K^{\ast}$ and it is sensitive
to all CP phases in $m_D$.
We give an example for the model in which J is determined 
by leptogenesis phase $\gamma_R$.
Suppose $ U_L$  is a real orthogonal matrix \cite{BMNR} as 
$U_L=O_{23} O_{12}$.
MNS matrix $K$ has the following form;
\bea
K= O_{23} O_{12} \left( \begin{array}{ccc}
 1 & 0 & 0  \\  
 0 & \cos\theta & \sin\theta \exp[-i \phi] \\
 0 & \sin\theta \exp[i \phi]  & \cos\theta \\ 
      \end{array} \right) P,
\eea
where $P$ is a diagonal phase matrix which is not relevant for 
$J$.
$\theta$ and $\phi$ are determined through the 
diagonalization of 
$ -U_L^{\dagger} m_D V_R \frac{1}{M} {V_R}^T {m_D}^T U_L^{\ast}$.
Therefore $\theta$ and $\phi$ do not depend on $U_L$ at all.
It depends on $m_2, m_3, \theta_R. \gamma_R$ besides $M_1$ and $M_2$.
Those four quantities can be determined from 
the heavy Majorana decay width, $\Gamma_1$ and
$\Gamma_2$ and two light neutrino masses scales $\Delta m_{atm.}$ and  
$\Delta m_{solar}$.
Taking $\theta_{23}=\pi/4$, $\theta_{12}=\pi/4$, we have;
\bea 
&& K=\left(\begin{array}{ccc}
      \frac{1}{\sqrt{2}} & \frac{\cos{\theta}}{\sqrt{2}} & 
      \frac{\sin{\theta} \exp[-i \phi]}{\sqrt{2}} \\
      -\frac{1}{2} & \frac{\cos\theta-\sqrt{2} 
      \sin \theta \exp[i \phi]}{2} & 
      \frac{\sin\theta \exp[-i \phi]+\sqrt{2} \cos \theta}{2}\\
      \frac{1}{2} & 
      -\frac{\cos\theta+\sqrt{2} \sin \theta \exp[i \phi]}{2}&  
      \frac{-\sin\theta \exp[-i \phi]+\sqrt{2} \cos \theta}{2}
      \\ \end{array} \right) P. 
\eea
It is easy to see $J \sim \sin \phi$. 
We can show there is a correlation between $\phi$ and $\gamma_R$.
If $\gamma_R$ vanishes, $m_{eff}$ becomes a real symmetric matrix.
Then, $\phi$ must vanish in the limit. On the other hand, if $\gamma_R$
does not vanish,
the sign of $\gamma_R$ determines the sign of lepton 
number asymmetry which in turn determines the excess of matter (anti-matter).
In our model, the sign of $J$ reflects the sign of $\gamma_R$.
We found the following correlation holds.
\bea
   B \rightarrow -B \Leftrightarrow L \rightarrow -L 
   \Leftrightarrow 
   \gamma_R \rightarrow -\gamma_R \Leftrightarrow \phi \rightarrow -\phi
  \Leftrightarrow
  J \rightarrow -J.
\eea 
In Fig.1 and Fig.2, 
we show the correlation $(\sin 2 \gamma_R, \sin \phi, x+y)$,
where $x= \frac{\Gamma_1 V^2}{{M_1}^2}$ and
$y=\frac{\Gamma_2 V^2}{{M_2}^2 }$,
by identifying the two light neutrino masses as 
$\sqrt{\Delta m _{atm.}^2}\sim  5.5 \times 10^{-2}$ eV and 
$\sqrt{\Delta m _{solar}^2}=(4 \sim 5) \times 10^{-3}$(LMA).
We take 
$\frac{M_1}{M_2}=0.1$.
The figures are obtained for fixed $(x+y) \times 10^{-2}$ (eV) and
varying $u=x-y$.
\newpage
\begin{figure}[htbp]
   \begin{minipage}{.48\textwidth}
   \hfill
	\includegraphics[scale=0.8]{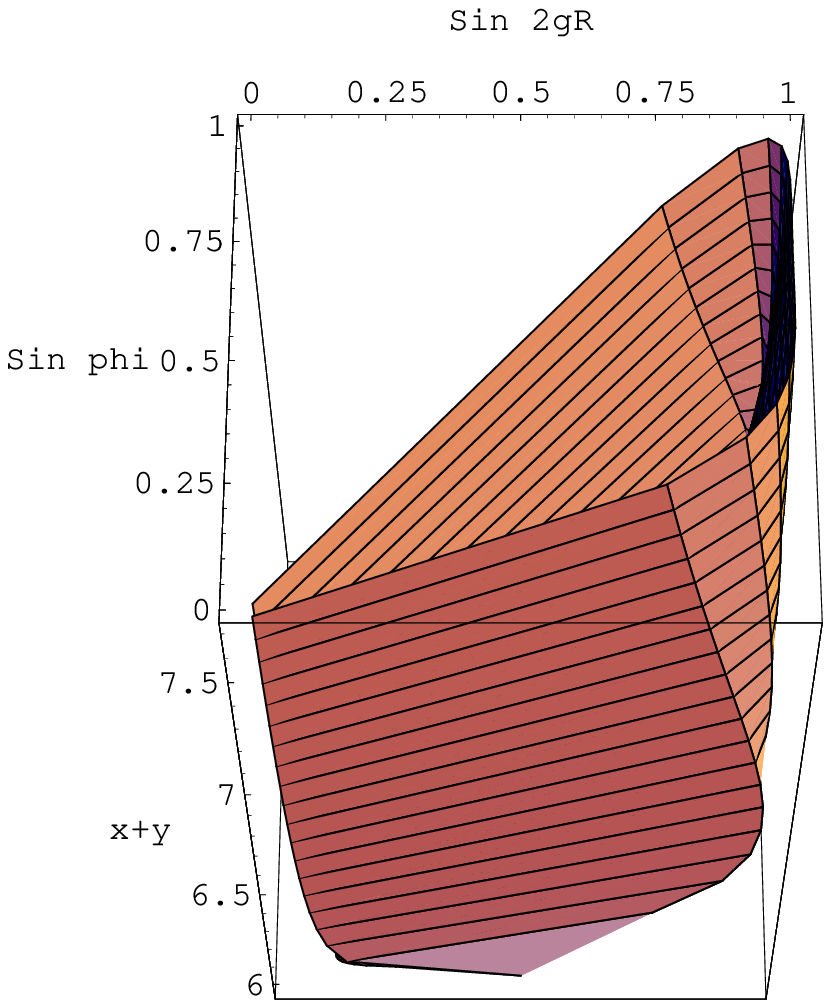}
	 \caption{\\
	 Correlation bet. leptogenesis CP phase and low energy CP phase;
	 $(\sin 2\gamma_R, \sin \phi, x+y)$. $0.06 < x+y < 0.075$ (eV).}
	   \end{minipage}
	     \begin{minipage}{.04\textwidth}
   \hfill
   \end{minipage}
   \begin{minipage}{.48\textwidth}
   \hfill
	\includegraphics[scale=0.7]{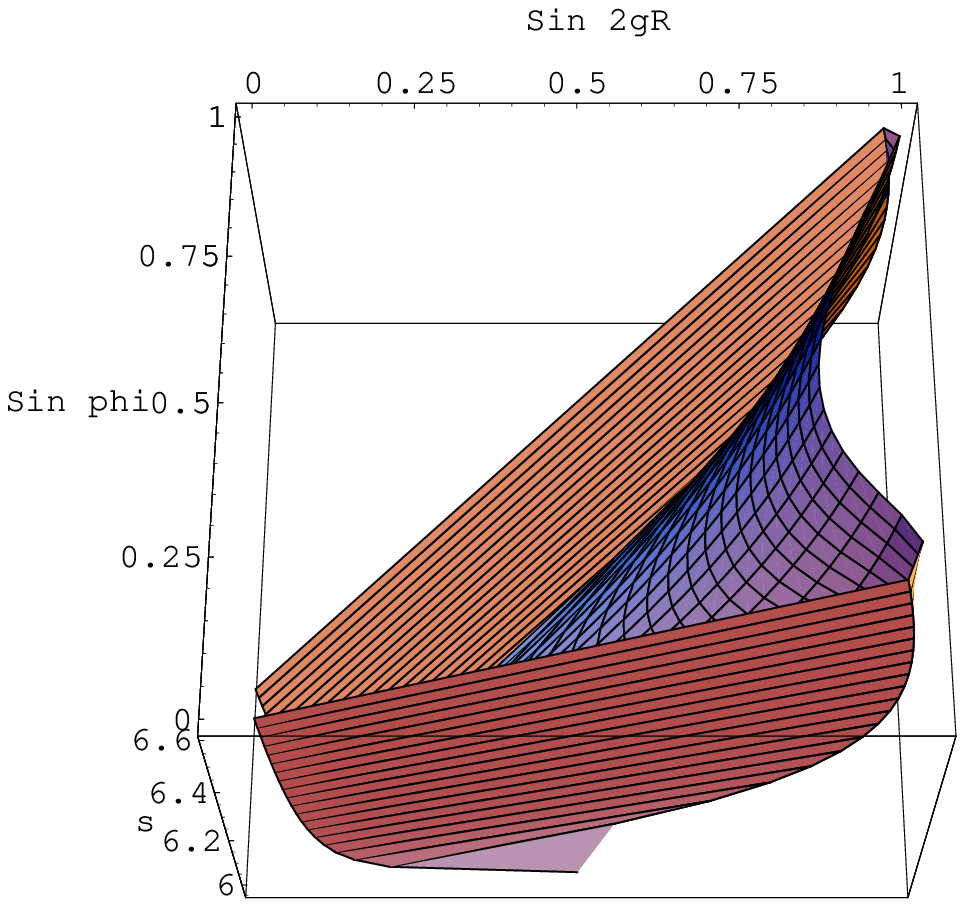}
	 \caption{\\
	 Correlation bet. leptogenesis CP phase and low energy CP phase;
	 $(\sin 2\gamma_R, \sin \phi, x+y)$. $0.06 <x+y <0.066$ (eV).}
	   \end{minipage}
\end{figure}	   
{\bf Summary}\\
\begin{itemize}
\item{leptogenesis phase ($\gamma_R$) certainly
affects the neutrino oscillation CP
violation through ($\phi$). However, if we measure $J$ only,
we can not distinguish the phase 
for the leptogenesis ($\phi$) from the other phases in $U_L$
because
only a certain combination of them appear.  The isolation must be
done using some other quantities, double $\beta$ decay etc \cite{EKKMT}.}
\item{ We show the correlation 
between CP violating phase for leptogenesis  and  
CP violating phase for neutrino oscillation for a specific choice of $U_L$.}
\end{itemize}
\ack{The works of T. M. and M. T. are supported by the Grand-in-Aid for Scientific Research of the MEXT, Japan, No.13640290 and No.12047220 respectively. 
S.K.K is supported by JSPS Invitation Fellowship (No.L02515)
and by BK21 program of the Ministry of
Education in Korea.
We thank H. So, A. Konaka, D. Chang, K. Hagiwara, 
J. Sato, T. Onogi, Y. Hayato, and O. Yasuda.}


\section*{References}
\begin{thebibliography}{99}
\bibitem{FuYa} M. Fukugita and T. Yanagida, Phys. Lett.{\bf B174}, 45(1986).
\bibitem{BMNR} G. Branco, T. Morozumi, B. Nobre and M. Rebelo,
Nucl. Phys. {\bf B 617}, 475 (2001).
\bibitem{EKKMT} T. Endoh, S. Kaneko, S. K. Kang, T. Morozumi, and 
M. Tanimoto, hep-ph 0209020.
\end{thebibliography}
\end{document}